\documentclass[conference]{IEEEtran}
\IEEEoverridecommandlockouts
\usepackage{cite}
\usepackage{amsmath,amssymb,amsfonts}
\usepackage{algorithm}
\usepackage{algpseudocode}
\usepackage{graphicx}
\usepackage{textcomp}
\usepackage{xcolor}
\usepackage{subfigure}
\usepackage{makecell}

\usepackage{threeparttable}
\usepackage{cuted}

\let\svthefootnote\thefootnote
\newcommand\freefootnote[1]{%
  \let\thefootnote\relax%
  \footnotetext{#1}%
  \let\thefootnote\svthefootnote%
}

\usepackage{amsmath}
\def\BibTeX{{\rm B\kern-.05em{\sc i\kern-.025em b}\kern-.08em
    T\kern-.1667em\lower.7ex\hbox{E}\kern-.125emX}}
\begin{document}

\title{Coexistence of resistive capacitive and virtual inductive effects in memristive devices  \\


}

\DeclareRobustCommand*{\IEEEauthorrefmark}[1]{%
  \raisebox{0pt}[0pt][0pt]{\textsuperscript{\footnotesize #1}}%
}

\author{\IEEEauthorblockN{Sahitya Yarragolla\IEEEauthorrefmark{1}*,
Torben Hemke\IEEEauthorrefmark{1}, Jan Trieschmann\IEEEauthorrefmark{2},\IEEEauthorrefmark{3} and Thomas Mussenbrock\IEEEauthorrefmark{1}}
\IEEEauthorrefmark{1}Chair   of   Applied   Electrodynamics   and Plasma   Technology,
Ruhr   University Bochum,
Bochum, Germany\\
\IEEEauthorrefmark{2}Theoretical Electrical Engineering, Faculty of Engineering, Kiel University, Kaiserstraße 2, 24143 Kiel, Germany\\
\IEEEauthorrefmark{3}Kiel Nano, Surface and Interface Science KiNSIS, Kiel University, Christian-Albrechts-Platz 4, 24118 Kiel, Germany\\

Email: *sahitya.yarragolla@rub.de}

\maketitle

\begin{abstract}
This paper examines the coexistence of resistive, capacitive, and inertia (virtual inductive) effects in memristive devices, focusing on ReRAM devices, specifically the interface-type or non-filamentary analog switching devices. A physics-inspired compact model is used to effectively capture the underlying mechanisms governing resistive switching in NbO$_{\rm x}$ and BiFeO$_{3}$ based on memristive devices. The model includes different capacitive components in metal-insulator-metal structures to simulate capacitive effects. Drift and diffusion of particles are modeled and correlated with particles' inertia within the system. Using the model, we obtain the \textit{I}-\textit{V} characteristics of both devices that show good agreement with experimental findings and the corresponding \textit{C}-\textit{V} characteristics. This model also replicates observed non-zero crossing hysteresis in perovskite-based devices. Additionally, the study examines how the reactance of the device changes in response to variations in the device area and length.

\end{abstract}


\section{Introduction}

\freefootnote{This work was funded by the Deutsche Forschungsgemeinschaft (DFG)—Project ID 434434223—SFB1461 and Project ID 439700144—Research Grant MU 2332/10-1 in the frame of Priority Program SPP 2253.}

Memristive devices have gained significant attention due to their potential applications in non-volatile memory \cite{Si2021}, neuromorphic computing \cite{Vasilopoulou2023}, and hardware security\cite{Rajendran2021}. They are characterized by their ability to remember past resistive states. The fundamental operation of a memristive device involves modulating resistance in response to an applied voltage or current, resulting in a change in memristance \cite{Chua2014}. Existing research on compact models for memristive devices has primarily focused on memristance, which is the memory of past electrical states. Numerous models have successfully captured the resistive effects, providing valuable insights into the behavior of these devices.

\begin{figure}[!t]
\centerline{\includegraphics[width=0.78\columnwidth]{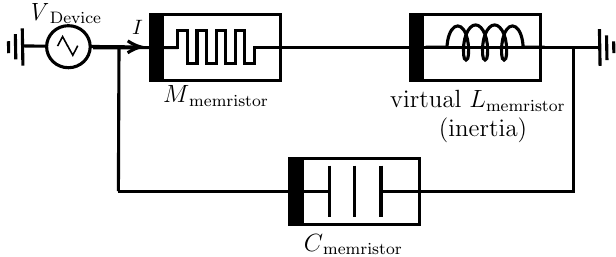}}
\caption{The coexistence of resistive, capacitive, and virtual inductive effects (inertial effects) in resistive switching devices.}
\label{fig:1}
\end{figure}

Understanding the complex electrical behavior of memristive devices, particularly in metal-insulator-metal (MIM) configurations, requires considering impedance that includes memristance $M_{\rm memristor}$, capacitance $C_{\rm memristor}$, and inertia (virtual inductive)  $L_{\rm memristor}$ as shown in Fig. \ref{fig:1}, rather than merely the change in resistance \cite{Qingjiang2014, Sun2020}. Traditional viewpoints often overlook the complex interplay of charged particles within the device by focusing solely on resistive effects. In MIM structures, the presence of charged particles introduces resistive, capacitive, and inertia effects that significantly influence the overall impedance. The charged particles contribute to capacitance by forming Schottky contact capacitance at the metal/oxide interface, oxide capacitance involving interactions between different charged particles, and tunnel barrier capacitance. Furthermore, the behavior of charged particles in the device can be compared to inductive effects, where their inertia affects the resistive switching behavior. In the analogy of electromechanics, change in the electric field that affects the particles' velocity also affects the particles' inertia. Therefore, referring to the inductive effects as inertia effects is more appropriate. By recognizing and including these capacitive and inertia elements in the broader concept of impedance, a more complete and precise model is created, providing better understanding of the complexities of memristive devices. 


This paper presents a new modification to an existing compact model that distinguishes it from conventional state-of-the-art models, especially for interface-type devices. The model is compact and includes the essential physical and chemical processes responsible for resistive switching \cite{Yarragolla2023}. Furthermore, it integrates the stochastic nature inherent in most memristive devices, providing a more realistic representation of device behavior. The framework was initially established as a cloud-in-cell (CIC) based model and has already demonstrated applicability for materials such as NbO$_{\rm x}$\cite{Yarragolla2022DBMD} and BiFeO$_{3}$\cite{Yarragolla2022BFO} devices. Our modification addresses a critical gap in existing models by incorporating capacitive and inertia effects. Subsequent sections provide a comprehensive explanation of the physical foundations of these effects. Furthermore, our research aims to investigate the performance of the modified model across different frequencies. It is recognized that including different effects introduces more intricate non-linear behaviors. This investigation contributes to refining our understanding of memristive devices and advancing the predictive capabilities of compact models in the context of emerging technologies.

\section{Simulation Approach}

            \begin{figure}[!t]
                \centering             \includegraphics[width=0.38\textwidth]{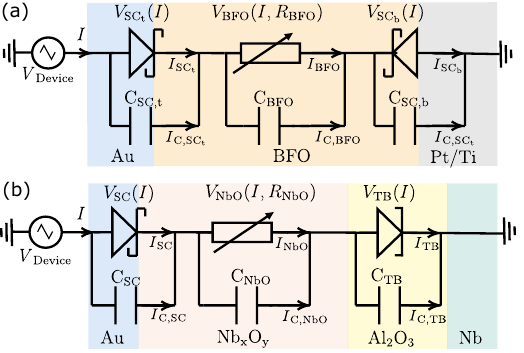}
                \caption{The modified equivalent circuits of (a) bismuth ferrite oxide memristive device \cite{Yarragolla2022BFO}  and (b) double barrier memristive device \cite{Yarragolla2022DBMD} with parallel capacitors across different layers. SC: Schottky contact, TB: tunnel barrier.}
                \label{fig:2}
            \end{figure}

The ReRAM devices, specifically the non-filamentary or interface-type or area-dependent devices such as bismuth ferrite oxide memristive devices (BFO), $\rm Au/BiFeO_{3}/Pt/Ti$ and double barrier memristive devices (DBMD), $\rm Au/Nb_{x}O_{y}/Al_{2}O_{3}/Nb$, consist of an oxide material sandwiched between two metal electrodes. Resistive switching (RS) in the BFO device shown in Fig. \ref{fig:2}(a) occurs by moving oxygen vacancies across the oxide layer and changing the interface properties. Similarly, the RS in DBMD in Fig. \ref{fig:2}(b) occurs via the movement of oxygen ions that change the interface properties. To model the RS in such devices that depend on ion/vacancy transport, a cloud-in-a-cell (CIC) \cite{Meyer2008} approach coupled with Kirchoff's laws for both devices, BFO and DBMD, has been demonstrated in \cite{Yarragolla2022BFO} and \cite{Yarragolla2022DBMD}, respectively. The proposed CIC-based model is used and extended further to incorporate capacitive and inertia effects.

\begin{figure}[!t]
                \centering             \includegraphics[width=0.37\textwidth]{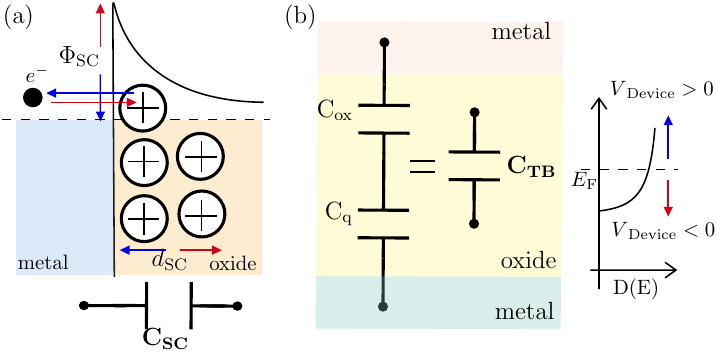}
                \caption{(a) Depletion layer width variation at LRS and HRS caused by trapping-de-trapping of electrons at the metal/insulator Schottky contact, influencing capacitance \cite{Yan2013}. (b) Tunnel barrier capacitance is represented as a series of combinations of oxide capacitance and quantum capacitance arising from voltage-induced changes in the density of states.}
                \label{fig:3}
            \end{figure}

When modeling capacitive effects in ReRAM devices, it is crucial to take a comprehensive approach that captures the various capacitive components inherent in these systems. This includes incorporating the Schottky contact capacitance, which arises from the metal/oxide interface, by considering the change in depletion region width of the metal/oxide interface ($d_{\rm SC}$) as shown in Fig. \ref{fig:3}(a) \cite{grundmann2015}. The rate at which the depletion width changes is given by $d_{\rm SC_{eff}}$, which depends on the internal state of the device ($q(t)$) and $\lambda_{d}$ is any random number between 0 and 1 taken in a way to match experimental results. The oxide capacitance encompasses the capacitance between various charged particles within the oxide layer and the oxide itself. The tunnel barrier capacitance is modeled as a series-connected capacitance, where both oxide and quantum capacitance dynamically adjust with changes across the tunnel barrier \cite{datta2005}. The quantum capacitance is taken as a change in the density of states $(D(E))$ with the change in $V_{\rm Device}$. The equations used for calculating the different capacitive components shown in Fig. \ref{fig:2} for both BFO and DBMD devices are given below:

\noindent 1. \textit{Schottky Contact \cite{grundmann2015}:}
\begin{subequations}
\begin{align}
        d_{\rm{SC}} = \sqrt{\frac{2\epsilon\left ( \Phi_{\rm{SC_{eff}}}-eV_{\textrm{SC}}-k_{\textrm{B}}T \right )}{en}}, \\
        d_{\rm{SC_{eff}}} = d_{\rm{SC}}(1 + \lambda_{d}\,q(t)), \hspace{0.5cm}{\rm and}\\ C_{\textrm{SC}}= \frac{\epsilon_{0}\epsilon_{\textrm{r}}A_{\textrm{d}}r_{\textrm{C}}}{d_{\rm{SC_{eff}}}}.\label{eq:newton-second-law}
\end{align}
\label{Eq:1}
\end{subequations}
Here $\epsilon$ is the permittivity of the oxide layer, $\Phi_{\rm SC_{eff}}$ is the effective value of the Schottky barrier height, $e$ is the elementary charge, $n$ is the defect density and $A_{\rm d}$ is the device area. $r_{\textrm{C}}$ is any value between 0 and 1, used to compensate for the unknown factors contributing to capacitive effects in oxides, $k_{B}$ is the Boltzmann constant, and $T$ is the temperature. 

\noindent 2. \textit{Oxide layer \cite{Yarragolla2022DBMD}:}
\begin{equation}
         d_{\rm{ox_{eff}}} = \frac{\sum_{i=1}^{N_{\rm p}}\left ( \bar{x}_{\rm i}-\bar{x}_{\rm interface}\right )}{N_{\rm p}} 
         \label{Eq:2}
\end{equation}
Here, $\bar{x}_{\rm i}$ is the position of $i_{\rm th}$ particle and $\bar{x}_{\rm interface}$ is the position of interface.

\noindent 3. \textit{Tunnel barrier \cite{datta2005}:}
\begin{subequations}\label{Eq:3}
\begin{align}
d_{\rm{TB_{eff}}} = d_{\rm{TB}}(1 + \lambda_{d}\,q(t))\\
        C_{\textrm{q}}= e^{2}D(E)\\
        C_{\textrm{TB}} = \frac{C_{\textrm{ox}}C_{\textrm{q}}}{C_{\textrm{ox}}+C_{\textrm{q}}}
\end{align}
\end{subequations}
Here, $d_{\rm{TB}}$ is the width of the tunnel barrier, $D(E)$ is the density of states, and $C_{\rm ox}$ is calculated using Eq. 1c. Due to the complexity of first-principle methods, for simplification, we assume a constant $D(E)(=4\times10^{5} \, {\rm eV^{-1}m^{-3}})$.

\begin{figure}[!b]
  \fbox{\parbox{0.95\columnwidth}{
    \footnotesize	
    \begin{algorithmic}[1]
      \State \textbf{Initialize:} Device parameters, grid ($N_{\rm grid}$)
      \For{each of $N_{\rm p}$ particles}
        \State Distribute particles, Interpolate charge to grid
      \EndFor
      \For{each step}
        \State Update Vdevice
        \For{each grid point}
          \State $\nabla^2\Phi = -\frac{\rho}{\varepsilon_0}$, $E = -\nabla\Phi$
        \EndFor
        \For{each particle}
          \State Calculate $\nu_D$ (Eq. \eqref{Eq:6}), $\bar{x}_{\rm i}=\bar{x}_{\rm i}+\Delta t \, \nu_D$
        \EndFor
        \State Calculate $d_{\text{eff}}, q(t)$ (Eqs. \eqref{Eq:1}-\eqref{Eq:3})
        \While{error $< 0.0001$}
          \State Assume $V_{\rm SC_{t}/TB} = V_{\text{device}}$
          \State Apply KVL, KCL and Ohm's law
          \State Calculate capacitive parameters (Eqs. 6-8)
          \State Calculate voltage drops ($V$), currents: $I_{\text{resistive}} + I_{\text{capacitive}}$ (Eqs. \eqref{Eq:7}-\eqref{Eq:10})
          \State error $= \frac{I_{\rm SC_{t}/TB} - I_{\rm SC_{b}}}{I_{\rm SC_{t}/TB}}$
          \State Update $V_{\rm SC_{t}/TB}$ based on the error
        \EndWhile
        \If{$N_{\rm steps}$ = $N_{\rm max}$}
          \State STOP
        \EndIf
      \EndFor
    \end{algorithmic}
  }}
  \caption{The pseudo code to implement the proposed CIC approach combined with Kirchoff laws, including capacitive and inertial effects.}
\label{fig:7}
\end{figure}

When investigating the coexistence of resistive, capacitive, and inertia effects in memristive devices, particularly ReRAMs, it is important to consider the inertia of charged particles, similar to the interconnection of inductors and resistors. This perspective is based on the electromechanical analogy \cite{Bloch1945}, where a simplified momentum conservation equation for positive or negative charges is given by,. 
\begin{equation}
m\frac{\mathrm{d} v_{\mathrm{D}}}{\mathrm{d} t}= \pm e E - m\gamma v_{\mathrm{D}}
 \label{Eq:4}
\end{equation}
After applying certain modifications to this equation, we arrive at the following equation,
\begin{equation}
V_{\rm ox}= \frac{m l_{\rm ox}}{e^2 n A_{\rm d}} \frac{\mathrm{d} I_{\rm ox}}{\mathrm{d} t} + \frac{m l_{\rm ox}}{e^2 n A_{\rm d}} \gamma I_{\rm ox}= L_{\rm ox} \frac{\mathrm{d} I_{\rm ox}}{\mathrm{d} t} + R_{\rm ox} I_{\rm ox}.
\label{Eq:5}
\end{equation}
A more detailed derivation and explanation can be found in \cite{Yarragolla2024}.
In this electrical equation, $L_{\rm SE}$ represents inductance, $R_{\rm SE}$ denotes resistance, and $m$ is the mass of the particle. The inertia of charged particles aligns with the inductive effects, encapsulating the particles' resistance to changes in motion. Conversely, the frictional force term finds an analogy in resistance, impeding the flow of charged particles. In the CIC approach, the drift velocity of the particle is calculated as follows:
\begin{equation}
     v_{\rm D} = \nu_{0} d \,\, {\rm exp}\left ( -\frac{{U}_{\rm A}}{k_{\rm B}T} \right ) \sinh\left (\frac{\left | z \right |edE}{k_{\rm B}T}  \right )
     \label{Eq:6}
\end{equation}
where $d$ is the lattice constant, $\nu_{0}$ is the phonon frequency, $z$ is the charge number of the ion, $E$ is the electric field and $U_{\rm A}$ is the activation energy. Calculating particle velocity at each time step and applying forces based on the input inherently integrates virtual inductance or inertia into the compact model. This computational approach mirrors the dynamics of charged particles and their response to external stimuli, effectively incorporating the effects of virtual inductance into the overall electrical behavior of the memristive device. 

Once the capacitive and inertia effects are incorporated, the currents across different regions are calculated as follows:

\noindent 1. \textit{Schottky contact:}
\begin{equation}
    I_{\rm SC} = A_{d}A^{*} T^{2}{\rm exp}\left \{ \frac{-\Phi_{\rm SC}}{k_{B}T} \right \}\left ( {\rm exp}\left \{ \frac{eV_{\rm SC}}{n_{\rm SC}k_{B}T} \right \} - 1\right ),
    \label{Eq:7}
\end{equation}
where, $n_{\rm SC}$ is the ideality factor and $A^{*}$ is the effective Richardson constant.

\noindent 2. \textit{Tunnel barrier:}
\begin{equation}
    \begin{split}
    I_{\rm TB} &= \frac{A_{d}e}{2\pi h\left (\beta d_{\rm TB}  \right )^{2}}\Biggl( \Phi_{\rm TB} \cdot {\rm exp}\left \{- {\rm A}\sqrt{\Phi_{\rm TB}}\right \} - \\
    &\left (\Phi_{\rm TB}+ e\left | V_{\rm TB} \right |  \right ) \cdot {\rm exp}\left \{- {\rm A} \sqrt{\Phi_{\rm TB}+ e\left | V_{\rm TB} \right | }\right \}\Biggr),
    \label{Eq:8}
    \end{split}
\end{equation}
where $\Phi_{\rm TB}$ and $d_{\rm TB}$ are the tunnel barrier height and length, and $A=\frac{4\pi\beta d\sqrt{2m} }{h}$. $\beta$, $m$, and $h$ are the correction factor, free electron mass, and the Planck constant.

\noindent 3. \textit{Oxide layer:}
\begin{equation}
    I_{\rm ox} = \sigma_{\rm ox} A_{\rm d}\frac{V_{\rm Ox}}{l_{\rm Ox}},
    \label{Eq:9}
\end{equation}
where $\sigma_{\rm ox}$ is the conductivity and $l_{\rm ox}$ is the length of the oxide.
\noindent 4. \textit{Capacitive current:}
\begin{equation}
    I_{\rm C} =  C\frac{\mathrm{d} V}{\mathrm{d} t}.
    \label{Eq:10}
\end{equation}

The pseudo code for implementing the proposed simulation method is given in Fig. \ref{fig:7}.


\section{Results and discussion}

\begin{figure}[!t]
\centering\includegraphics[width=0.4\textwidth]{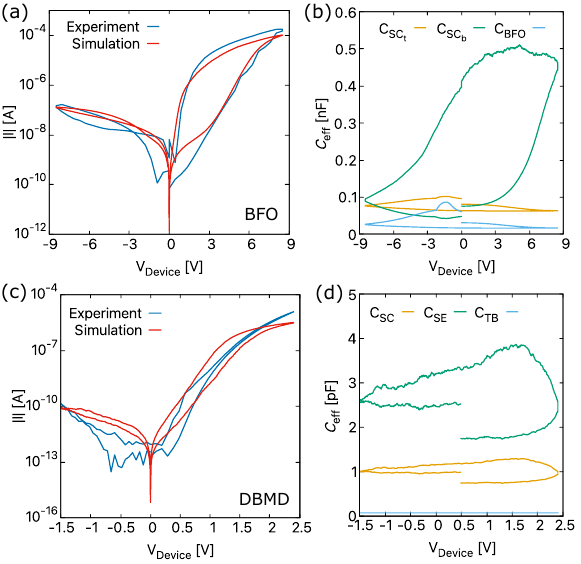}
\caption{The \textit{I}-\textit{V} characteristics with capacitive and inertial effects and \textit{C}-\textit{V} characteristics of (a)-(b) BFO\cite{Du2018, Yarragolla2022BFO}  and (c)-(d) DBMD devices\cite{Yarragolla2022DBMD}.}
\label{fig:4}
\end{figure}

The \textit{I}-\textit{V} characteristics of BFO and DBMD devices are shown in Fig. \ref{fig:4}(a) and Fig. \ref{fig:4}(c), respectively. The calculated \textit{I}-\textit{V} curves match quite well with the experimental findings. Moreover, for BFO, the non-zero crossing point is also quite well mimicked by incorporating different effects. This shows that it is very important to consider these effects in computational models to be closer to the physics and replicate the results more precisely. At very low voltages, almost close to 0\,V, the calculated and the experimental curves are not so similar. This is attributed to the noise content in the experimental findings. 

The change in capacitance with voltage (\textit{C}-\textit{V} curves) across different regions of the devices are shown in Fig. \ref{fig:4}(b) and Fig. \ref{fig:4}(d). As observed from both \textit{C}-\textit{V} curves, the change in capacitance across the Schottky contacts majorly contributes to the change in the capacitance in a device over other capacitive components. Compared with BFO, the change in capacitance is much less in DBMD. The \textit{I}-\textit{V} curves of DBMD are not much affected by the capacitance and therefore, we observe a zero crossing in \textit{I}-\textit{V} curve. 

\begin{figure}[!t]
\centering\includegraphics[width=0.47\textwidth]{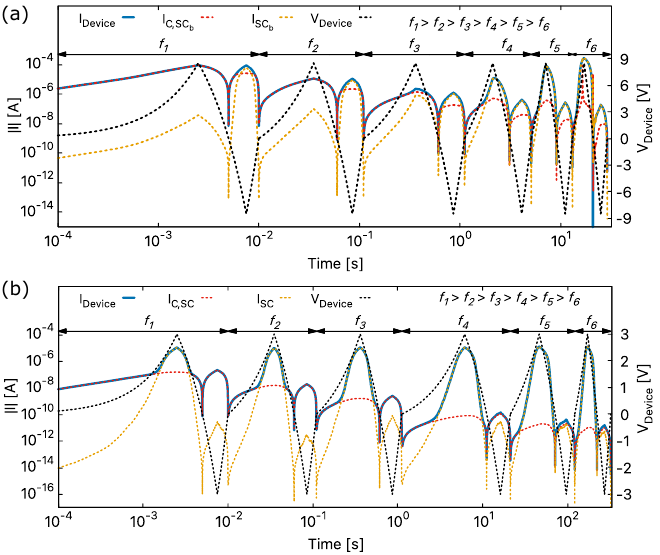}
\caption{The change total current, resistive, and capacitive currents in (a) BFO and (b) DBMD devices for a voltage of constant amplitude and varying frequency.}
\label{fig:5}
\end{figure}

Analyzing the response of a memristive device to different voltage frequencies is crucial to understanding its complex electrical behavior. Investigating changes in current provides insights into capacitive and inertia effects, with different frequencies affecting the interaction of resistive, capacitive, and inertia components. This exploration is necessary to optimize device performance and accurately model its behavior in practical applications. Fig. \ref{fig:5} shows the change in overall memristive device current, Schottky current calculated by Eq. \eqref{Eq:7}, and capacitive current with changing frequency. These plots are helpful to understand the contribution of each current at different frequencies to the overall current. For the BFO device, a triangular voltage of 8.5\,V is used; for DBMD, a similar voltage of 3\,V is used. As observed in Fig. \ref{fig:5}, as the frequency decreases, the change in capacitive current decreases, and the contribution of resistive current across the Schottky contact increases. At increased (decreased) frequencies, the capacitive reactance decreases (increases) due to its reciprocal relationship with frequency $(X_{\rm C}=1/2\pi fC)$. This results in a rapid increase (decrease) in capacitive current. Almost no phase shift is observed since the capacitance is in the nF range $(\phi= {\rm tan}^{-1}(1/\omega RC))$.  

\begin{figure}[!t]
\centering\includegraphics[width=0.4\textwidth]{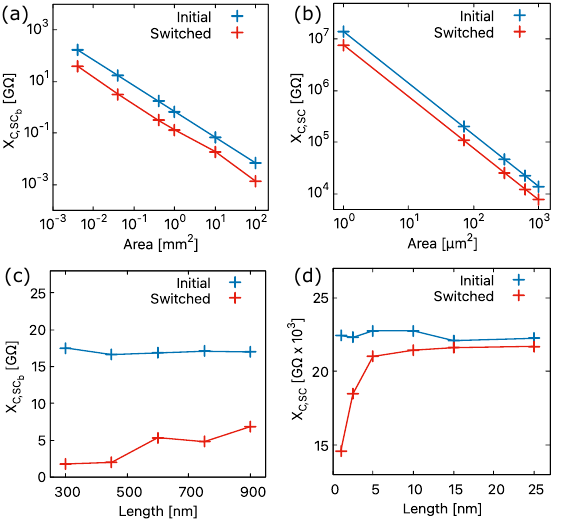}
\caption{The initial (HRS) and switched (LRS) capacitive reactance for change in device area and length of the oxide layer. (a) and (c) BFO device. (b) and (d) DBMD device. }
\label{fig:6}
\end{figure}

To further understand the contribution of capacitive effects in Fig.\ref{fig:6}, we plotted the initial reactance, i.e., the reactance at the high resistance state and the switched reactance, that is, the reactance when the device is switched to a low resistance state. The values are calculated for BFO for a frequency of 0.125\,Hz, 8.5\,V  and DBMD of 0.01\,Hz, 3\,V. The figures illustrate the transition from initial to switched reactance in the memristive device, highlighting the influence of varying device area and oxide layer length. At high resistance, the initial capacitive reactance is sensitive to device area and oxide length changes. Specifically, an increase in device area increases capacitance, while an increase in oxide length correlates with an increase in Schottky depletion layer width. Together, these changes contribute to an increase in capacitive reactance. In contrast, the switched reactance at low resistance is characterized by different responses to changes in area and oxide layer length. The decrease in device area results in a decrease in capacitance, accompanied by a contraction of the Schottky depletion layer width as the oxide length increases. As a result, the switched reactance decreases.

\section{Conclusion}

In conclusion, this paper explores resistive, capacitive, and inertia (virtual inductive effects) coexistence in memristive devices, specifically non-filamentary ReRAM devices. The compact model integrates capacitive and inertia, improving its accuracy in capturing the intricacies of resistive switching in NbO$_{x}$ and BiFeO$_{3}$ based devices. Capacitive effects due to charged particles within the oxide layer and particle inertia effects significantly improve the model's predictive capabilities. The modified model faithfully reproduces the observed non-zero crossing hysteresis and agrees well with experimental current-voltage characteristics. This investigation highlights the need to model virtual effects to represent memristive devices' physics accurately. The simulation results highlight the significant impact of capacitive and inertia elements on the overall device behavior, providing insight into their interplay and relevance in advanced memristive systems.



\bibliographystyle{IEEEtran}
\bibliography{references}  

\begin{thebibliography}{10}
\providecommand{\url}[1]{#1}
\csname url@samestyle\endcsname
\providecommand{\newblock}{\relax}
\providecommand{\bibinfo}[2]{#2}
\providecommand{\BIBentrySTDinterwordspacing}{\spaceskip=0pt\relax}
\providecommand{\BIBentryALTinterwordstretchfactor}{4}
\providecommand{\BIBentryALTinterwordspacing}{\spaceskip=\fontdimen2\font plus
\BIBentryALTinterwordstretchfactor\fontdimen3\font minus \fontdimen4\font\relax}
\providecommand{\BIBforeignlanguage}[2]{{%
\expandafter\ifx\csname l@#1\endcsname\relax
\typeout{** WARNING: IEEEtran.bst: No hyphenation pattern has been}%
\typeout{** loaded for the language `#1'. Using the pattern for}%
\typeout{** the default language instead.}%
\else
\language=\csname l@#1\endcsname
\fi
#2}}
\providecommand{\BIBdecl}{\relax}
\BIBdecl

\bibitem{Si2021}
\BIBentryALTinterwordspacing
M.~Si, H.-Y. Cheng, T.~Ando, G.~Hu, and P.~D. Ye, ``Overview and outlook of emerging non-volatile memories,'' \emph{MRS Bulletin}, vol.~46, no.~10, pp. 946--958, Oct 2021. [Online]. Available: \url{https://doi.org/10.1557/s43577-021-00204-2}
\BIBentrySTDinterwordspacing

\bibitem{Vasilopoulou2023}
\BIBentryALTinterwordspacing
M.~Vasilopoulou, A.~R.~b. Mohd~Yusoff, Y.~Chai, M.-A. Kourtis, T.~Matsushima, N.~Gasparini, R.~Du, F.~Gao, M.~K. Nazeeruddin, T.~D. Anthopoulos, and Y.-Y. Noh, ``Neuromorphic computing based on halide perovskites,'' \emph{Nature Electronics}, vol.~6, no.~12, pp. 949--962, Dec 2023. [Online]. Available: \url{https://doi.org/10.1038/s41928-023-01082-z}
\BIBentrySTDinterwordspacing

\bibitem{Rajendran2021}
\BIBentryALTinterwordspacing
G.~Rajendran, W.~Banerjee, A.~Chattopadhyay, and M.~M.~S. Aly, ``Application of resistive random access memory in hardware security: A review,'' \emph{Advanced Electronic Materials}, vol.~7, no.~12, p. 2100536, 2021. [Online]. Available: \url{https://onlinelibrary.wiley.com/doi/abs/10.1002/aelm.202100536}
\BIBentrySTDinterwordspacing

\bibitem{Chua2014}
\BIBentryALTinterwordspacing
L.~Chua, ``If it’s pinched it’s a memristor,'' \emph{Semiconductor Science and Technology}, vol.~29, no.~10, p. 104001, sep 2014. [Online]. Available: \url{https://dx.doi.org/10.1088/0268-1242/29/10/104001}
\BIBentrySTDinterwordspacing

\bibitem{Qingjiang2014}
\BIBentryALTinterwordspacing
L.~Qingjiang, A.~Khiat, I.~Salaoru, C.~Papavassiliou, X.~Hui, and T.~Prodromakis, ``Memory impedance in tio2 based metal-insulator-metal devices,'' \emph{Scientific Reports}, vol.~4, no.~1, p. 4522, Mar 2014. [Online]. Available: \url{https://doi.org/10.1038/srep04522}
\BIBentrySTDinterwordspacing

\bibitem{Sun2020}
\BIBentryALTinterwordspacing
B.~Sun, M.~Xiao, G.~Zhou, Z.~Ren, Y.~Zhou, and Y.~Wu, ``Non–zero-crossing current-voltage hysteresis behavior in memristive system,'' \emph{Materials Today Advances}, vol.~6, p. 100056, 2020. [Online]. Available: \url{https://www.sciencedirect.com/science/article/pii/S2590049820300035}
\BIBentrySTDinterwordspacing

\bibitem{Yarragolla2023}
S.~Yarragolla, T.~Hemke, and T.~Mussenbrock, ``A generic compact and stochastic model for non-filamentary analog resistive switching devices,'' in \emph{2023 12th International Conference on Modern Circuits and Systems Technologies (MOCAST)}, 2023, pp. 1--4.

\bibitem{Yarragolla2022DBMD}
\BIBentryALTinterwordspacing
S.~Yarragolla, T.~Hemke, J.~Trieschmann, F.~Zahari, H.~Kohlstedt, and T.~Mussenbrock, ``Stochastic behavior of an interface-based memristive device,'' \emph{Journal of Applied Physics}, vol. 131, no.~13, p. 134304, Apr 2022. [Online]. Available: \url{https://doi.org/10.1063/5.0084085}
\BIBentrySTDinterwordspacing

\bibitem{Yarragolla2022BFO}
\BIBentryALTinterwordspacing
S.~Yarragolla, N.~Du, T.~Hemke, X.~Zhao, Z.~Chen, I.~Polian, and T.~Mussenbrock, ``Physics inspired compact modelling of bifeo3 based memristors,'' \emph{Scientific Reports}, vol.~12, no.~1, p. 20490, Nov 2022. [Online]. Available: \url{https://doi.org/10.1038/s41598-022-24439-4}
\BIBentrySTDinterwordspacing

\bibitem{Meyer2008}
R.~Meyer, L.~Schloss, J.~Brewer, R.~Lambertson, W.~Kinney, J.~Sanchez, and D.~Rinerson, ``{Oxide dual-layer memory element for scalable non-volatile cross-point memory technology},'' in \emph{Proceedings - 9th Annual Non-Volatile Memory Technology Symposium, NVMTS}, 2008.

\bibitem{grundmann2015}
\BIBentryALTinterwordspacing
M.~Grundmann, \emph{The Physics of Semiconductors: An Introduction Including Nanophysics and Applications}, ser. Graduate Texts in Physics.\hskip 1em plus 0.5em minus 0.4em\relax Springer International Publishing, 2015. [Online]. Available: \url{https://books.google.de/books?id=VEdECwAAQBAJ}
\BIBentrySTDinterwordspacing

\bibitem{datta2005}
S.~Datta, \emph{Quantum Transport: Atom to Transistor}.\hskip 1em plus 0.5em minus 0.4em\relax Cambridge University Press, 2005.

\bibitem{Yan2013}
\BIBentryALTinterwordspacing
Z.~B. Yan and J.-M. Liu, ``Coexistence of high performance resistance and capacitance memory based on multilayered metal-oxide structures,'' \emph{Scientific Reports}, vol.~3, no.~1, p. 2482, Aug 2013. [Online]. Available: \url{https://doi.org/10.1038/srep02482}
\BIBentrySTDinterwordspacing

\bibitem{Bloch1945}
\BIBentryALTinterwordspacing
A.~Bloch, ``Electromechanical analogies and their use for the analysis of mechanical and electromechanical systems,'' \emph{Journal of the Institution of Electrical Engineers - Part I: General}, vol.~92, pp. 157--169, 1945. [Online]. Available: \url{https://api.semanticscholar.org/CorpusID:110010613}
\BIBentrySTDinterwordspacing

\bibitem{Yarragolla2024}
S.~Yarragolla, T.~Hemke, J.~Trieschmann, and T.~Mussenbrock, ``Non-zero crossing current-voltage characteristics of interface-type resistive switching devices,'' 2024.

\bibitem{Du2018}
\BIBentryALTinterwordspacing
N.~Du, N.~Manjunath, Y.~Li, S.~Menzel, E.~Linn, R.~Waser, T.~You, D.~B\"urger, I.~Skorupa, D.~Walczyk, C.~Walczyk, O.~G. Schmidt, and H.~Schmidt, ``Field-driven hopping transport of oxygen vacancies in memristive oxide switches with interface-mediated resistive switching,'' \emph{Phys. Rev. Applied}, vol.~10, p. 054025, Nov 2018. [Online]. Available: \url{https://link.aps.org/doi/10.1103/PhysRevApplied.10.054025}
\BIBentrySTDinterwordspacing

\end{thebibliography}

\end{document}